\documentclass[%
 amsmath,amssymb,
 aps,
longbibliography,
]{revtex4-1}

\usepackage{dcolumn}
\usepackage{bm}
\usepackage[colorlinks=true,allcolors=blue]{hyperref}

\usepackage{graphicx}
\graphicspath{{Figures/}} 
\usepackage{epstopdf}
\usepackage{color}

\usepackage{cleveref}

\begin{document}

\title{Nanoconfined ionic liquids: disentangling electrostatic \& viscous forces}

\author{Romain Lhermerout}
\author{Susan Perkin}%
 \email{susan.perkin@chem.ox.ac.uk}
\affiliation{Department of Chemistry, Physical and Theoretical Chemistry Laboratory, University of Oxford, Oxford OX1 3QZ, United Kingdom}

\date{\today}

\begin{abstract}
Recent reports of surface forces across nanoconfined ionic liquids have revealed the existence of an anomalously long-ranged interaction apparently of electrostatic origin. Ionic liquids are viscous and therefore it is important to inspect rigorously whether the observed repulsive forces are indeed equilibrium forces or, rather, arise from the viscous force during drainage of the fluid between two confining surfaces. In this paper we present our direct measurements of surface forces between mica sheets approaching in the ionic liquid $[\mathrm{C_2 C_1 Im}][\mathrm{NTf_2}]$, exploring three orders of magnitude in approach velocity. Trajectories are systematically fitted by solving the equation of motion, allowing us to disentangle the viscous and equilibrium contributions. First, we find that the drainage obeys classical hydrodynamics with a negative slip boundary condition in the range of the structural force, implying that a $\mathrm{nm}$-thick portion of the liquid in the vicinity of the solid surface is composed of ordered molecules that do not contribute to the flow. Secondly, we show that a long-range static force must indeed be invoked, in addition to the viscous force, in order to describe the data quantitatively. This equilibrium interaction decays exponentially and with decay length is in agreement with the screening length reported for the same system in previous studies. In those studies the decay was simply checked to be independent of velocity and measured at low approach rate, rather than explicitly taking account of viscous effects: we explain why this gives indistinguishable outcomes for the screening length by noting that the viscous force is linear to very good approximation over a wide range of distances.

\end{abstract}

\maketitle

\section{Introduction}
\label{Introduction}

The Surface Force Apparatus (SFA) or Balance (SFB) was originally designed by Tabor and Winterton to measure Van der Waals interactions between macroscopic bodies surrounded by air or vacuum \cite{Israelachvili2011a,Tabor1969a}. Subsequent extensions by Israelachvili led to measurement of surface interactions across liquids \cite{Israelachvili1978a} and many varieties of soft matter. Aqueous solutions of ions have been used extensively to study electrostatic interactions \cite{Israelachvili1978a,Horn1988a,Pashley1981a,Pashley1984a}; in particular this allowed validation of the DLVO theory of colloidal stability in dilute electrolytes. In these works, the viscosity and velocity were small enough to safely neglect viscous effects and to consider only equilibrium forces.

Separately, Chan \& Horn took advantage of the capabilities of the SFA to investigate the viscous force in nanoconfined liquids \cite{Chan1985a}. They focused on purely dissipative forces by using apolar liquids, and showed that drainage is quantitatively well-described by classical hydrodynamics down to the nanoscale. For the liquids used, that completely wet mica surfaces, the best agreement was obtained by considering a non-slip boundary condition a few nanometers away from the solid surface in the liquid phase, thereafter reported as negative slip. From this observation emerged the picture of a stopped layer, namely a thin film in the vicinity of the solid surface that is part of the liquid phase but does not contribute to the fluid flow. Since the pioneering work of Chan \& Horn, many studies looked at the hydrodynamic boundary condition with the SFA, generalizing the analysis to many systems \cite{Israelachvili1986a,Horn1988b,Israelachvili1989a,Horn1989b}, some of them presenting partial slip \cite{Vinogradova1995a,Bocquet2010a} or elastohydrodynamic deformation \cite{Leroy2011a,Leroy2012a,Villey2013a,Wang2015a,Wang2017a,Wang2017b}.

Very recently ionic liquids (ILs) have been the focus of intense study because of their exceptional properties (negligible volatility, nonflammability, high ionic conductivity, wide electrochemical and thermal windows) which make them extremely promising for a range of engineering and electrochemical applications (energy storage, lubrication, catalysis, electrowetting, etc.) \cite{Fedorov2014a,Hayes2015a}. Forces of apparently electrostatic origin have been reported with very substantially longer range than anticipated based on existing theories, leading to an active debate in the community as to their interpretation \cite{Gebbie2013a,Perkin2013a,Gebbie2013b,EspinosaMarzal2014b,Cheng2015a,Lee2015a,Gebbie2015a,Smith2016a,Hjalmarsson2017a,Comtet:2017aa,Gebbie2017a,Lee2017a,Lee2017b}. To measure such long-range forces, these studies used the same protocol as for aqueous solutions: using low enough velocities for the dynamical effects to be negligible, i.e. checking that the measured force was independent of velocity over a range of slow velocities. However ionic liquids have dynamic viscosity typically 100 times greater than that of water, consequently one can estimate the viscous force to be of the same order of magnitude as the reported electrostatic forces at the approach velocities employed. It is therefore necessary to formally disentangle the equilibrium and viscous contribution to the measured forces \cite{Horn1989a}. More generally, it is of interest to ask whether classical hydrodynamics applies to nanoconfined ionic liquids, and whether a slip or stick boundary condition applies. Indeed, ionic liquids are composed solely of charged molecules with strong electrostatic interactions that could potentially modify the drainage rate because of coupled electro-diffusio-osmotic mechanisms. In particular, when the liquid is confined at a scale within the range of a long-range electrostatic interaction, electroneutrality is broken and one can expect substantial electrokinetic effects \cite{Levine1975a,Bocquet2010a}.

Thus a systematic study is needed in order to tackle these dynamical aspects. This paper is dedicated to answering the following questions. \textit{Does classical hydrodynamics apply for nanoconfined ILs and what is the slip boundary condition? When considering quantitatively the viscous force during drainage, does ILs still exhibit a long-range electrostatic interaction?} We report SFB measurements exploring 3 orders of magnitude in velocity, and the equations of motion combining viscous and electrostatic forces have been systematically solved to analyze the data. Such approach allowed us to disentangle static and dynamic contributions to the total force when the ionic liquid, here $[\mathrm{C_2 C_1 Im}][\mathrm{NTf_2}]$, is confined between approaching mica surfaces. We found first no significant deviation from Reynolds theory of fluid drainage, and that classical hydrodynamics holds down to the nanoscale with a negative slip length of $4.2\pm0.7~\mathrm{nm}$, i.e. a layer of a few molecules that are stopped in the vicinity of the solid surface. We then show that an additional equilibrium, long-range exponentially decaying contribution has to be invoked to describe the total force quantitatively. The decay length of this interaction is found to be $8.0\pm0.5~\mathrm{nm}$, in agreement with the previously reported values for this system. We finally come back to the interpretation of previous studies, and explain why ignoring a non-negligible viscous force turns out nonetheless to provide a correct evaluation of the range of the electrostatic force.

The manuscript is organized as follows. The experimental set-up and the protocol of analysis is presented in section~\ref{Materials_and_Methods}; the measurement at the highest velocity is shown in section~\ref{High_velocities} in order to characterize the purely viscous response; data at very low velocities are then used in section~\ref{Very_low_velocities} to reveal the equilibrium electrostatic repulsion; and an intermediate velocity is finally explored in section~\ref{Intermediate_velocities} to comment on previous analyses.

\section{Materials and Methods}
\label{Materials_and_Methods}

\begin{figure}[h!]
\includegraphics{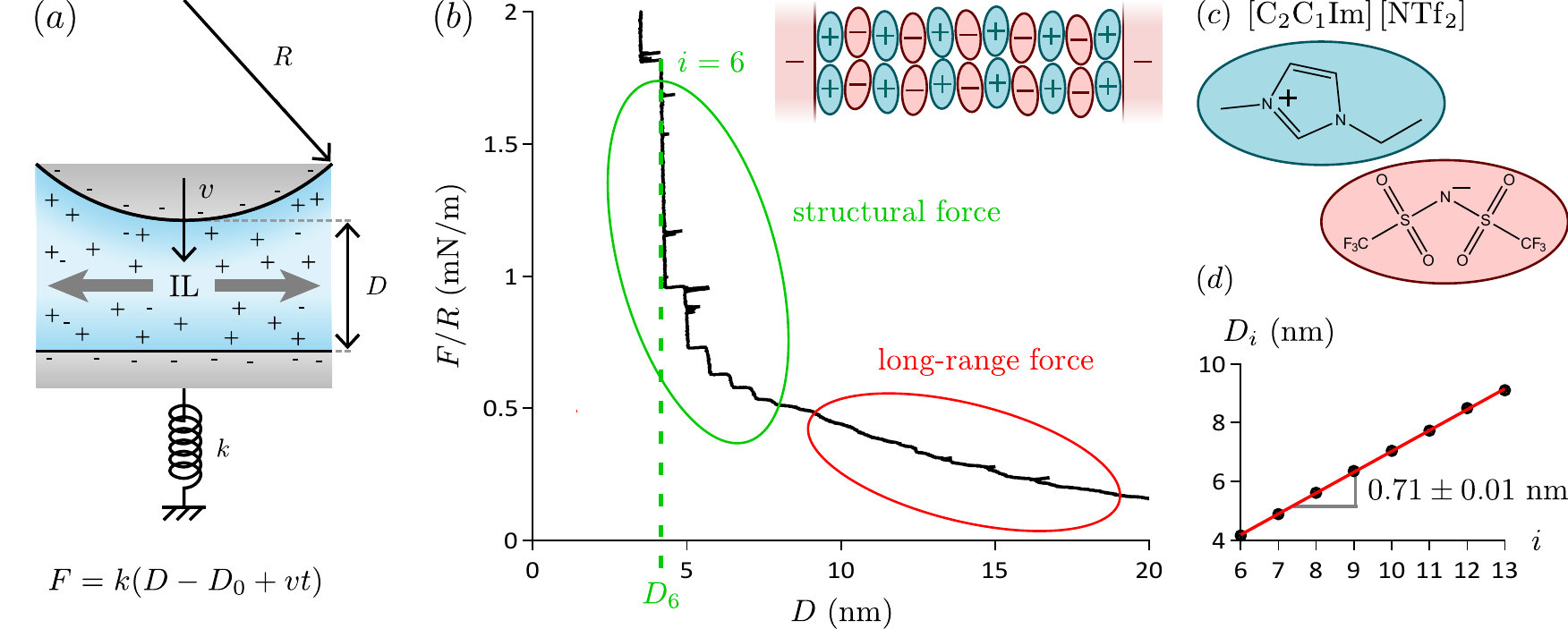}
\caption{ a) Schematic showing the principle of SFB measurements. b) Force profile measured during approach of two mica surfaces separated by $[\mathrm{C_2 C_1 Im}][\mathrm{NTf_2}]$ at $v=0.075\pm 0.010\mathrm{~nm/s}$. At short distances, below ca. 7-8 nm one observes a structural force with characteristic steps due to successive squeeze-out of ordered layers (also called solvation force). The configuration with $i=6$ cations along the gap is indicated by the green dashed line and is illustrated in the inset. At larger distances a long-range force is found, and the aim of this paper is to determine the static and dynamic contributions to it. c) Chemical structure of $[\mathrm{C_2 C_1 Im}][\mathrm{NTf_2}]$. d) Location of the layers in the structural force region: the slope gives the average layer thickness.}
\label{Fig1}
\end{figure}

The principle of the SFB is illustrated in \crefformat{figure}{Figure~#2#1{(a)}#3}\cref{Fig1}\crefformat{figure}{Fig~#2#1#3}, and explained in detail elsewhere \cite{Israelachvili2011a,Klein1998a,Perkin2006a}. Briefly, muscovite mica is cleaved to produce micrometer-thick facets that are molecularly smooth, which are then backsilvered and glued onto cylindrical glass lenses with an epoxy resin (EPON~1004, Shell Chemicals). Two lenses made with mica of the same thickness are arranged in a cross-cylinder geometry, and separated by a liquid bridge. The liquid used is \textcolor{black}{1-ethyl-3-methylimidazolium \textit{bis}[(trifluoromethane)sulfonyl]imide} $[\mathrm{C_2 C_1 Im}][\mathrm{NTf_2}]$ \textcolor{black}{ (Iolitec, 99\%)}, which chemical structure is given in \crefformat{figure}{Figure~#2#1{(c)}#3}\cref{Fig1}\crefformat{figure}{Fig~#2#1#3}, notably because it is one of the less viscous ILs (dynamic viscosity $\eta=37~\mathrm{mPa.s}$ at $25^\circ \mathrm{C}$ \cite{Yu2012a}). The liquid is dried in a Schlenck line at $60^\circ \mathrm{C}$ and $2\times 10^{-2}~\mathrm{mbar}$
 for 10 hours, and injected in the chamber just after. Typically one hour is waited before measurements for the apparatus to be in thermal equilibrium with the room regulated to $25^\circ \mathrm{C}$.

The top lens is translated with a piezo-electric tube (Ferroperm PZ 29) that allows a very smooth and linear motion over a typical range of $200~\mathrm{nm}$, at velocity $v$ controlled between $0.07~\mathrm{nm/s}$ (limited by drifts) and $100~\mathrm{nm/s}$ (limited by the camera frame rate). White light interferences between the silver mirrors are observed though a spectrometer, and the so-called FECO (Fringes of Equal Chromatic Order) pattern is recorded. Knowing the refractive index of $[\mathrm{C_2 C_1 Im}][\mathrm{NTf_2}]$ (1.42251 at $25^\circ \mathrm{C}$ \cite{Tariq2009a}), it is possible to extract the radius of curvature $R=8.85~\mathrm{mm}$ and the liquid thickness $D$ \cite{Israelachvili1972a}. The bottom lens is mounted on a spring (spring constant $k=125~\mathrm{N/m}$), deflection of which allows us to deduce the normal force between the two surfaces.

Since the last publication of our group reporting force profiles \cite{Smith2017a}, our setup has been improved to significantly reduce the noise level on the signal $D(t)$. First, a QImaging Retiga R6 camera provides a high resolution ($2688 \times 2200~\mathrm{px}^2$), a better signal to noise ratio and a FPS of $14~\mathrm{Hz}$. Secondly the apex position of the FECO fringes is extracted from a local parabolic fitting. Thirdly, a box made of expanded polystyrene sheets surrounds the SFB in order to damp the vibrations coming from air. These modifications allows us to achieve a RMS noise of $0.02~\mathrm{nm}$ and so very clean signals (see \crefformat{figure}{Figure~#2#1{(b)}#3}\cref{Fig1}\crefformat{figure}{Fig~#2#1#3}) that are ideal for fitting weak long-range forces.

In this version of the SFB, only the distance $D$ between the lenses is measured directly, and the total force is deduced from it with the relation $F(t) = k\left(D(t) - D_0+vt\right)$, where $D_0$ is the starting distance, which is supposed to be large enough to have $F(0)=0$. Here the velocity~$v$ (resp. force~$F$) has been taken positive for approaching motion (resp for repulsive interaction). A difficulty is that the velocity $v$ of the top lens is not precisely known in advance, as there is a 10\% fluctuation from run to run, and a calibration of such uncertainty prevents from reliably identifying weak long-range forces. The method commonly used to solve this problem consists in fitting linearly the $D(t)$ trajectory at distances large enough to have negligible long-range forces, so that the fitted slope corresponds to $v$. This method is excellent when dealing with aqueous solutions and velocities such that the viscous force is below the limit of detection of the apparatus, but does not necessarily apply for viscous liquids and velocities such that viscous force is measurable. In this paper, we follow the approach of Chan \& Horn by solving the equation of motion of the bottom lens and fitting the $D(t)$ trajectories \cite{Chan1985a}.

Neglecting the inertia of the bottom lens, the equation of motion in the frame of the lab is simply the balance of the total force $F$ exerted by the top lens across the liquid and of the spring restoring force. For our system, we model the total force by the superposition of two forces: an equilibrium, electrostatic force, and a dynamic, viscous force. A strong assumption is that these two contributions are supposed to be totally independent, i.e. no electrokinetic effect is included.

To describe the viscous force, we suppose that the nanoconfined IL behaves like a charge-less, Newtonian liquid in an incompressible flow between undeformable solids. In the lubrication approximation ($D \ll R$), the viscous force is simply given by Reynolds formula. As suggested by Chan \& Horn, we use a modified version of the Reynolds formula, with hydrodynamic zero shifted by the distance $b$ in the liquid (see the viscous term in equation~\ref{eq_motion}). This "negative slip length" can be seen as the presence of a "stopped layer" on each solid surface, i.e. liquid films of thickness $b/2$ that do not contribute to the drainage flow, on top of which a no-slip boundary condition applies.

The electrostatic interaction is modelled by an exponential function of decay length~$\lambda_\mathrm{s}$ and amplitude~$2\pi R A$, where~$2\pi R$ is a geometrical factor coming from the Derjaguin approximation. Such an exponential law is expected to hold in the context of the linear Debye-H\"uckel theory for diluted electrolytes while imposing a constant potential boundary condition between two symmetric planar surfaces \cite{Israelachvili2011a}, which is clearly not the current situation. Although the origin of the exponentially decaying force in ionic liquids and concentrated electrolytes is still under discussion, it appears clear that experiments to date agree on the empirical functional form \cite{Gebbie2017a,Lee2017a,Lee2017b}. Thus the equation of motion finally reads:

\begin{equation}
k\left(D(t)-D_0+vt\right)=F(t)=\textcolor{black}{-}\frac{6\pi\eta R^2 \dot{D}}{D-b}+2\pi R Ae^{-D/\lambda_\mathrm{s}}.
\label{eq_motion}
\end{equation}

Note that the structural force, observed experimentally for this system (\crefformat{figure}{Figure~#2#1{(b)}#3}\cref{Fig1}\crefformat{figure}{Fig~#2#1#3}), as well as the Van der Waals interaction are not included in the model for the sake of simplicity. Each of these is short ranged compared to the viscous and electrostatic interactions in pure ionic liquid and neglecting them does not alter the fitting of data in this work.

An analytic solution only exists in the purely viscous case ($A=0$) \cite{Chan1985a}, therefore the equation of motion is integrated numerically, using the Runge-Kutta method implemented in~\verb!C++!. Four parameters are known: the radius $R$ and the initial distances $D_0$ are measured, the spring constant $k$ has been calibrated prior to the experiment and the viscosity~$\eta$ is known from literature. The four other parameters are manually tuned to fit the data: first the velocity $v$ is fitted at large distances, secondly the hydrodynamic zero $b$, the screening length $\lambda_\mathrm{s}$ and the amplitude $A$ are fitted at small scales, and finally all the parameters are refined iteratively. For a single trajectory, the parameters $b$, $\lambda_\mathrm{s}$ and $A$ are interdependent, but as we will see later the exploration of 3 decades of velocities allows to disentangle the role of each parameter. In short, high velocities provide first the determination of the hydrodynamic zero $b$ only, from there the screening length $\lambda_\mathrm{s}$ and the amplitude $A$ can be fitted unequivocally at very low velocities.

After this fitting procedure, when the imposed motion~$D_0-vt$ and so the spring deflection are precisely known, the total force can be calculated and separated into viscous and electrostatic contributions.

For the sake of clarity, only data obtained at four different approach velocities are shown in the following section. In practice, very many approaches have been carried out experimentally to ensure the reproducibility of the observations and the robustness of the interpretations.

\section{Results and Discussion}
\label{Results_and_Discussion}

The force profile obtained upon approach at the lowest velocity $v=0.075\pm0.010\textrm{~nm/s}$ is shown in \crefformat{figure}{Figure~#2#1{(b)}#3}\cref{Fig1}\crefformat{figure}{Fig~#2#1#3}. Here the rescaled force $F/R$ is used, because according to the Derjaguin approximation this quantity is proportional to the interaction energy acting between the same system in the plane-plane geometry, and is convenient to compare measurements performed in different geometries.

At short distances ($D<10~\mathrm{nm}$ typically), one observes a so-called structural force, with characteristic steps. This phenomenon is known to be the signature of the ordering of the molecules in layers in the confined film. A jump-in corresponds to the sudden squeeze-out of a layer, which -- for ionic liquids -- must consist of equal numbers of positive and negative ions to preserve electroneutrality. Here we show only the repulsive steps which contribute to the overall oscillatory structural force law, and we only probe the range down to $~4\mathrm{nm}$: the present experiment was optimised for weak and long range forces (using a piezo with short range). Other studies have probed the full oscillations in the past \cite{Smith2017a,Smith2015a}. Our present measurements are consistent with the previous observations in terms of the positions and amplitude at the maxima. The closest (small $D$) maximum explored here, indicated by the green dashed line in \crefformat{figure}{Figure~#2#1{(b)}#3}\cref{Fig1}\crefformat{figure}{Fig~#2#1#3}, corresponds to the sixth layer and - with our now-improved resolution as described above - we are able to probe a further seven layers. That is to say, steps in the force profile corresponding to molecular layers (structural forces) are measured out to 13 layers in total. The remaining imperfections in the signal are horizontal "spikes", that are due to mechanical vibrations that are not completely damped. In \crefformat{figure}{Figure~#2#1{(d)}#3}\cref{Fig1}\crefformat{figure}{Fig~#2#1#3} the location of the maxima $D_i$ are plotted against the layer index $i$. The linearity of the curve obtained is very good, illustrating the regularity of the layers. The slope is the average layer thickness, and is found to be $0.71\pm 0.01~\mathrm{nm}$. This is close to a previous AFM determination and to the predicted ion pair diameter of $0.75\pm 0.01~\mathrm{nm}$ \cite{Hayes2010a}. Our value is also slightly smaller than the $0.78\pm0.01\textrm{~nm}$ previously reported \cite{Smith2015a}, although in that case the layer thickness was extracted from the locations of minima and from deepest layers.

At larger distances ($D>10~\mathrm{nm}$ typically) we also observe a long-range force, with no distinguishable layers, decaying much slower than the structural force. This monotonic force will be our focus for the remainder, in which we consider its equilibrium and dynamic contributions.

\subsection{High velocities: characterization of the dynamical response}
\label{High_velocities}

\begin{figure}[h!]
\includegraphics{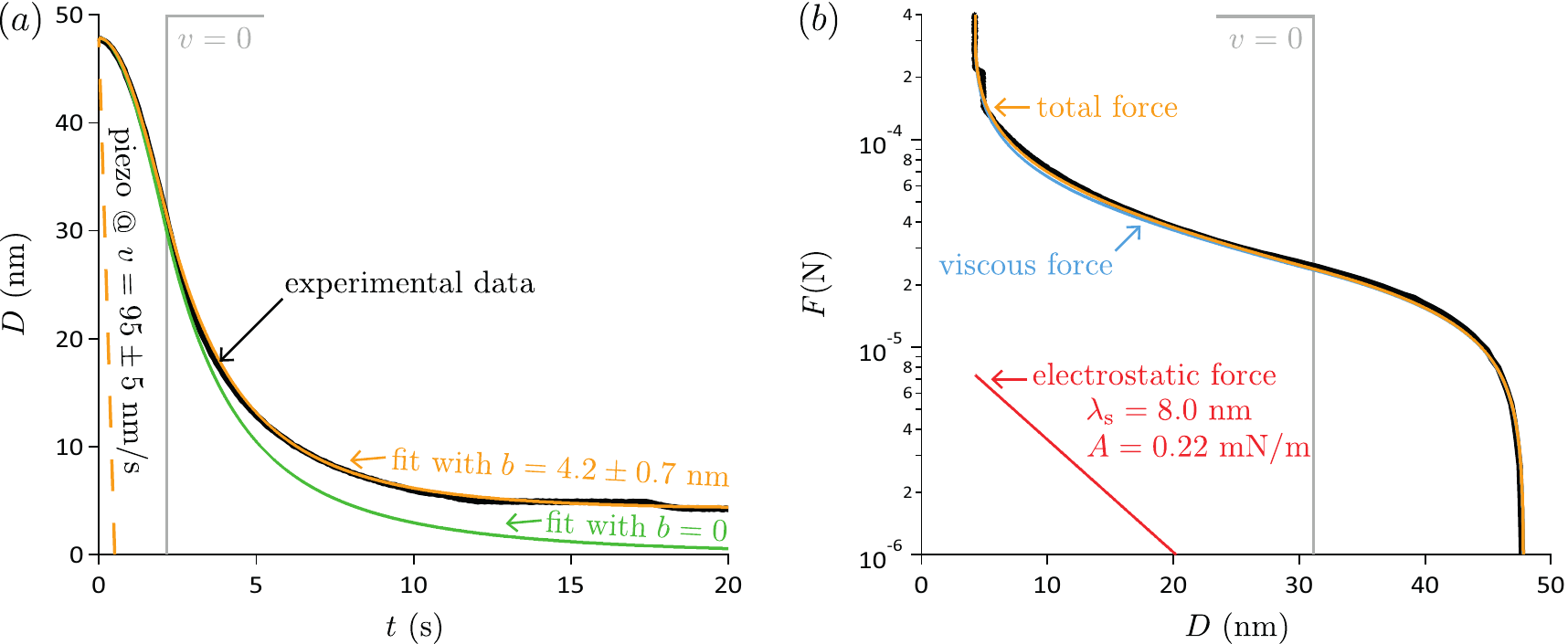}
\caption{(a)~Trajectory recorded when the top lens is moving at $v=95\pm 5\textrm{~nm/s}$ (piezo motion indicated by the dashed line in orange). The black curve is the experimental signal, the orange (resp. green) curve is the fit by the model with $b=4.2\pm 0.7\textrm{~nm}$ (resp. $b=0$), $\lambda_\mathrm{s}=8.0\textrm{~nm}$ and $A=0.22\textrm{~mN/m}$. These values of $\lambda_\mathrm{s}$ and $A$ are justified by fits at lower velocity as described later; as noted in the text a fit with $A=0$ would be indistinguishable at this velocity. (b)~Same measurements, but with data presented in terms of force, in semi-log scales. The black curve is obtained from the difference between the black curve and the orange dashed line in (a). The total force (in orange), viscous contribution (in blue) and electrostatic contribution (in red) are deduced from the fit. In both graphs, the vertical line in gray indicates the time and distance at which the piezo reaches its maximum deformation and the top lens stops.}
\label{Fig2}
\end{figure}

\crefformat{figure}{Figure~#2#1{(a)}#3}\cref{Fig2}\crefformat{figure}, shows in black the trajectory recorded when the top lens is moved at a high velocity. At such high velocity the viscous force is expected to dominate and the data could be fitted by assuming a purely viscous force ($A=0$), with $v=95\pm 5~\mathrm{nm/s}$ and $b=4.2\pm0.7~\mathrm{nm}$. In this regime, the viscous damping is very clearly significant: while the top lens is moving at $v=95\pm 5~\mathrm{nm/s}$, the relative velocity $\dot{D}$ between the lenses does not exceed $12~\mathrm{nm/s}$. Because of this huge viscous force, the piezo-electric tube reaches its maximum range and stops before the contact between the two lenses, at a time indicated by the vertical gray line. Nevertheless, the approach towards contact continues after, because of the elastic energy stored in the spring. The stop of the piezo motion at a known time is taken into account when solving the equation of motion, so that this experimental detail is not a problem to interpret the data.

The beginning of the trajectory ($D\gtrsim20~\mathrm{nm}$) is fitted with a single parameter, namely the velocity $v$. Indeed, the initial curvature is well-captured by the model, whatever value is chosen for the hydrodynamic zero, $b$. The fitted velocity is found to be close within 10\% to the value expected by independent calibrations in air. We thus have a good agreement with the Reynolds formula ($b=0$) whenever the confinement is moderate. At stronger confinement ($7~\mathrm{nm}\lesssim D\lesssim20~\mathrm{nm}$), a simple no-slip boundary condition on the solid surface ($b=0$) clearly does not allow a good fit to the data (green curve in \crefformat{figure}{Figure~#2#1{(a)}#3}\cref{Fig2}\crefformat{figure}), and the best fit is obtained with $b=4.2\pm0.7~\mathrm{nm}$ (orange curve in \crefformat{figure}{Figure~#2#1{(a)}#3}\cref{Fig2}\crefformat{figure}), which corresponds to the location of the layer $i=6\pm 1$. Note that the model doesn't fit the steps-like shape at $D\lesssim7~\mathrm{nm}$, because it doesn't include the structural and Van der Waals forces.

Interestingly, we distinguish only three layers, which locations correspond to the three deepest layer over the nine layers seen at $v=0.075\pm 0.010\mathrm{~nm/s}$ (\crefformat{figure}{Figure~#2#1{(b)}#3}\cref{Fig1}\crefformat{figure}{Fig~#2#1#3}). Such decrease of the range of the structural force with the approach velocity has already been observed for apolar molecules \cite{Bureau2008a,Bureau2012a}. To interpret this phenomenon, the authors proposed a scenario in which the rapidly squeeze-out film is quenched in a metastable disordered state. This aspect is beyond the scope of the present work but it is of interest to note that this phenomenon can also be observed with ionic liquids.

Our determination of a negative slip length of $[\mathrm{C_2 C_1 Im}][\mathrm{NTf_2}]$ on mica is consistent with recent measurements made by Garcia et al. \cite{Garcia2017a}. They use a modified SFA, which is able to superpose a tiny oscillation to the steady motion, and so allows to distinguish the conservative (in-phase) and dissipative (out-of-phase) responses. For $[\mathrm{C_4 C_1 Im}][\mathrm{PF_6}]$ between Pyrex or Platinum surfaces, they concluded that the steady flow obeys classical hydrodynamics with a no-slip boundary condition on top of a few nanometers thick film of ordered molecules. Intriguingly, they found that the oscillating flow simultaneously obeys to classical hydrodynamics with no-slip boundary condition close to the solid surface. This observation suggests that the ordered layers of molecules can participate or not to the flow, depending on the nature of the excitation.

Furthermore, it is worth noting that the extracted value of $b=4.2\pm0.7~\mathrm{nm}$ seems to match with the liquid thickness below which a stick-slip, solid-like response is \textcolor{black}{observed} in friction measurements \cite{Perkin2010,Smith2013a}. In these experiments, the top lens is moved laterally and the liquid is sheared at strains comparable with our steady approaches in the normal direction. These two independent observations are thus consistent with the picture of ordered layers of molecules that don't flow when subjected to a large scale excitation.

As we will see in section~\ref{Very_low_velocities}, our system clearly exhibits an electrostatic interaction at very low velocity. Anticipating this result, we analyzed the data at high velocity while incorporating the static contribution determined at very low velocity (i.e. with $A=0.22~\mathrm{mN/m}$ and $\lambda_\mathrm{s}=8.0~\mathrm{nm}$). At such high velocity, the electrostatic force is much smaller than the total force (see \crefformat{figure}{Figure~#2#1{(b)}#3}\cref{Fig2}\crefformat{figure}), and considering a purely viscous response is a very good approximation. However, given we know that there is an electrostatic force, we could have expected some electrokinetic effects, especially when probing confinement within the range $\lambda_\mathrm{s}=8.0~\mathrm{nm}$ of the electrostatic force. In the end we found no measurable deviation from classical hydrodynamics, considering the IL as a neutral, Newtonian liquid of constant, bulk viscosity.

Finally, one can ask why we do not observe elastohydrodynamic effects \cite{Leroy2011a,Leroy2012a,Villey2013a,Wang2015a,Wang2017a,Wang2017b}. Indeed, although the model assumes undeformable solids, the materials used experimentally are not infinitely stiff. We have two symmetric lenses which are stacks composed of $3.6083~\mathrm{\mu m}$ of mica / $\sim 50~\mathrm{nm}$ of silver / $\sim 10~\mathrm{\mu m}$ of EPON supported by a centimetric glass cylinder. We can estimate the order of magnitude of the indentation $\delta$ of the solid, by simply considering a rigid sphere approaching a soft, semi-infinite medium. For undeformable solids and the simple case of a no-slip boundary condition, the sphere exerts the Reynolds force $6\pi\eta R^2 \dot{D}/ D$ on a radius of action roughly given by the range $\sqrt{RD}$ of the pressure distribution. Taking these scaling as first estimates for the soft case, we can write Hooke's law \cite{Leroy2011a}:

\begin{equation}
\frac{\delta}{\sqrt{RD}} \sim \frac{1}{E^\star}\frac{6\pi\eta R^2 \dot{D}/ D}{\pi RD},
\label{eq_hooke}
\end{equation}

\noindent
where $E^\star$ is the effective elastic modulus, given by $E/(1-\nu^2)$ for a semi-infinite medium, where $E$ is the Young's modulus and $\nu$ the Poisson's ratio. As expected, the indentation increases with the viscosity, the velocity and the softness of the substrate. The indentation also increases when the confinement increases, and we can define a typical size $D_\mathrm{c}$ such that $\delta \sim D$. This lengthscale corresponds to a situation where the actual liquid film thickness is doubled in comparison to the rigid case, and provides a scale above which elasticity cannot be neglected. From equation~\ref{eq_hooke} we finally get the scaling:

\begin{equation}
D_\mathrm{c} \sim \left[\frac{\eta R^{3/2} \dot{D}}{E^\star}\right]^{2/5}
\label{eq_Dc}
\end{equation}

\noindent
Our lenses are not semi-infinite, but are stratified with typical Young's moduli of $\sim 200~\mathrm{GPa}$ for the mica, $\sim 50~\mathrm{GPa}$ for the glass or the silver, $1~\mathrm{GPa}$ for the EPON \cite{Wang2015a}. In first approximation, we take the effective modulus $E^\star\sim10~\mathrm{GPa}$ extracted from JKR tests on similar systems \cite{Horn1987a}, even if it has been shown that effective considerations in contact deformation cannot be extended to elastohydrodynamic deformation \cite{Wang2015a,Wang2017b}. For the maximum relative velocity $\dot{D}\sim10~\mathrm{nm/s}$ reached in our experiment, we finally obtain $D_\mathrm{c}\sim 1~ \mathrm{nm}$. Qualitatively, this very rough estimation means that elastic deformation could play a role when the two surfaces are closer than a few $\mathrm{nm}$. Additional observations support the scenario of negligible elastohydrodynamics. First, no significant deformation of the lenses is observed when looking at the FECO pattern. Secondly, similar experiments have been performed with the same system and parameters, except that the liquid is a PDMS polymer melt \cite{Horn1988b,Horn1989a}, and reported no significant elastohydrodynamic effect. Finally, we reproduced these measurements in our set-up with a similar PDMS polymer melt (Xiameter PMX 200 Silicone Fluid $50~\mathrm{cSt}$), leading to the same conclusion.

\subsection{Very low velocities: the equilibrium contribution revealed}
\label{Very_low_velocities}

\begin{figure}[h!]
\includegraphics{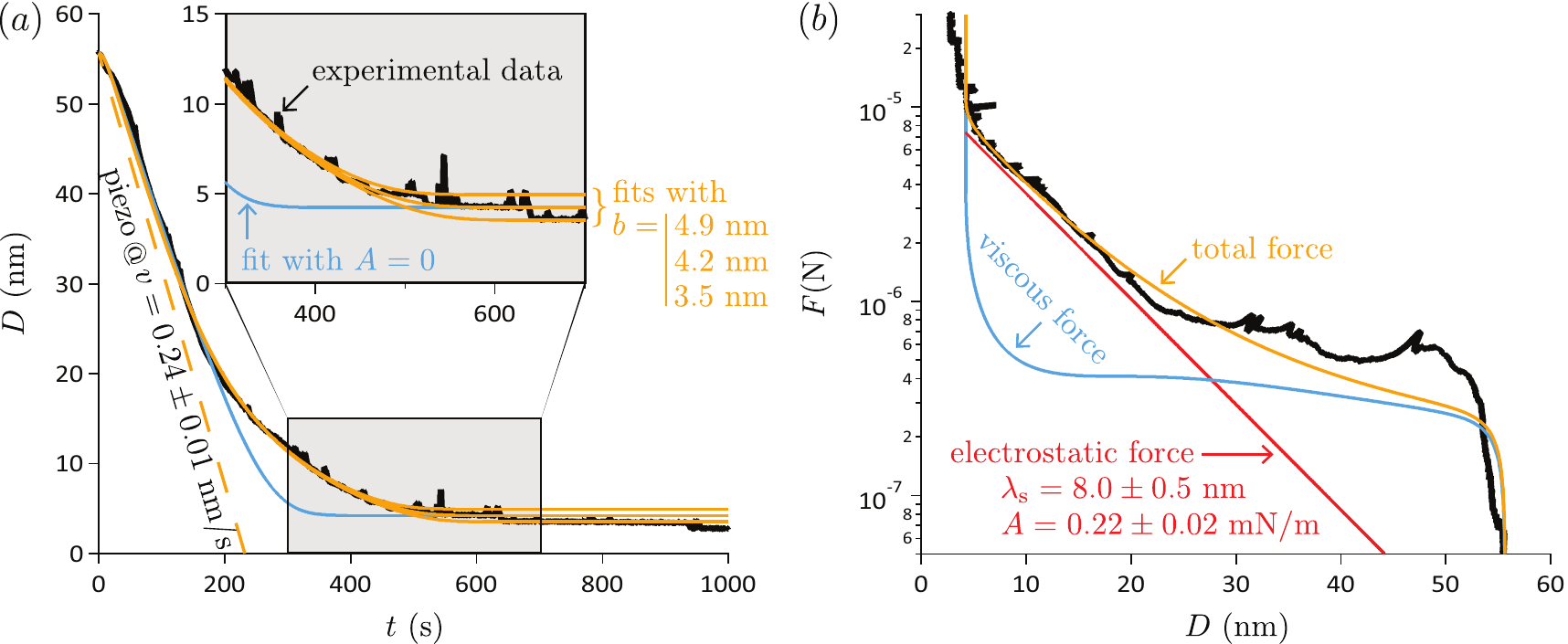}
\caption{(a)~Trajectory recorded when the top lens is moving at $v=0.24\pm 0.01\textrm{~nm/s}$ (piezo motion indicated by the dashed line in orange). The black curve is the experimental signal, the blue curve is an attempt to fit with a single, viscous force ($A=0$), and the orange curve is the fit by the model with $\lambda_\mathrm{s}=8.0\pm0.5\textrm{~nm}$ and $A=0.22\pm0.02\textrm{~mN/m}$ and three different values $b\in\{3.5~\mathrm{nm};4.2~\mathrm{nm};4.9~\mathrm{nm}\}$ (zoom in inset). (b)~Same measurements, but with data presented in terms of force, in semi-log scales. The black curve is obtained from the difference between the black curve and the orange dashed line in (a). The total force (in orange), viscous contribution (when fitting with non-zero A; in blue) and electrostatic contribution (in red) are deduced from the fit.}
\label{Fig3}
\end{figure}

In \crefformat{figure}{Figure~#2#1{(a)}#3}\cref{Fig3}\crefformat{figure}, are shown in black the measurements performed at a very low velocity. The hydrodynamic zero is kept to its value $b=4.2~\mathrm{nm}$ extracted before, but in this regime the trajectory cannot be fitted with a purely viscous force (see blue line which corresponds to $A=0$). At intermediate distances ($7~\mathrm{nm}\lesssim D\lesssim\textcolor{black}{25}~\mathrm{nm}$), a significant repulsion is not captured by a purely viscous model. On the contrary, the trajectory can be very well-fitted when considering a static, exponentially decaying contribution, interpreted as an electrostatic repulsion (see the orange curve). The best fit is obtained with a velocity $v=0.24\pm 0.01~\textrm{nm/s}$ (see the orange dashed line), a decay length $\lambda_\mathrm{s}=8.0\pm 0.5~\mathrm{nm}$ and an amplitude $A=0.22\pm 0.02~\mathrm{mN/m}$. The trajectories computed for three different values of $b\in\{3.5~\mathrm{nm};4.2~\mathrm{nm};4.9~\mathrm{nm}\}$ (orange curves in \crefformat{figure}{Figure~#2#1{(a)}#3}\cref{Fig3}\crefformat{figure}, and zoom in inset) show that the hydrodynamic zero $b$ modifies very weakly the parameters fitted for the static contribution.

The same data are presented in terms of force in \crefformat{figure}{Figure~#2#1{(b)}#3}\cref{Fig3}\crefformat{figure}, on a semi-log scale. At large distances ($D\gtrsim 25~\mathrm{nm}$), the main contribution is the viscous force. The fit in this region is not perfect, as the semi-log scales emphasizes the imperfections due to thermal drifts. At intermediate distances ($7~\mathrm{nm}\lesssim D\lesssim 25~\mathrm{nm}$), the total force varies over one order of magnitude with a shape compatible with an exponential behavior. The electrostatic force is the dominant contribution, as the viscous force is one order of magnitude smaller. Additionally, the shape of the viscous force is significantly different than the total force, and it is clear that it is not possible to fit the data with a purely viscous force, for example by playing with the hydrodynamic zero $b$ that basically translates the curve along the horizontal axis. Lastly, the region of very strong confinement ($D\lesssim 7~\mathrm{nm}$) is not fitted by the model; unsurprising because the model does not include the structural or Van der Waals forces.

To have a good determination of the static contribution, it is important to highlight that the initial distance should be large enough: $D_0 \gg \lambda_\mathrm{s}$. If not, the values of $v$ and $\lambda_\mathrm{s}$ cannot be determined independently from the fit. This is a well-known experimental requirement, which has to be fulfilled even when neglecting the viscous effects. 

Finally, the screening length $\lambda_\mathrm{s}=8.0\pm 0.5~\mathrm{nm}$ obtained here is in reasonable agreement with the values previously reported for this system: $6.7\pm 0.1~\mathrm{nm}$ \cite{Gebbie2015a} and $7.1\pm 0.7~\mathrm{nm}$ \cite{Smith2016a}. This might at first appear surprising given that, in the earlier studies, the viscous contribution was not formally taken into account. In the next section we rationalise why the electrostatic decay lengths match nonetheless.

\subsection{Intermediate velocities: why ignoring viscosity \textcolor{black}{can be} taking it into account}
\label{Intermediate_velocities}

\begin{figure}[h!]
\includegraphics{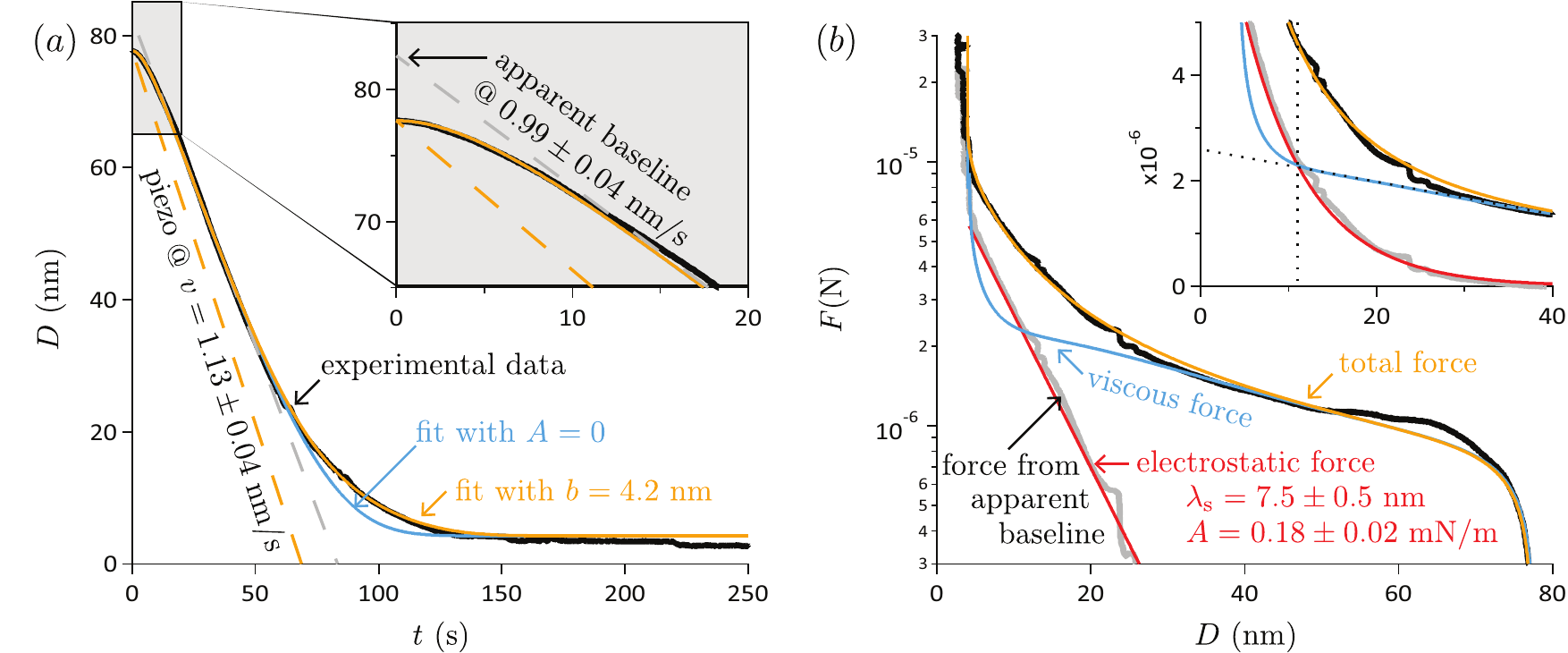}
\caption{(a)~Trajectory recorded when the top lens is moving at $v=1.13\pm 0.04\textrm{~nm/s}$ (piezo motion indicated by the dashed line in orange). The black curve is the experimental signal, the blue curve is an attempt to fit with a single, viscous force ($A=0$) with $b=4.2\textrm{~nm}$, and the orange curve is the fit by the model with $\lambda_\mathrm{s}=7.5\pm0.5\textrm{~nm}$ and $A=0.18\pm0.02\textrm{~mN/m}$. The gray dashed line is a linear fit a large distances, which doesn't take into account the strong curvature observed initially (zoom in inset) and provides the velocity $0.99\pm0.04\textrm{~nm/s}$. (b)~Same measurements, but with data presented in terms of force, in semi-log scales. The black curve is obtained from the difference between the black curve and the orange dashed line in (a). The total force (in orange), viscous contribution (in blue) and electrostatic contribution (in red) are deduced from the fit with non-zero $A$, whereas the force profile in gray line is computed from the gray dashed line in (a). Inset: same data plotted in linear scales. The vertical dashed line in gray indicates the intersection between the viscous and electrostatic contributions, and the second gray dashed line is a guide for the eyes to illustrate the linearity of the viscous component at large distances.}
\label{Fig4}
\end{figure}

\crefformat{figure}{Figure~#2#1{(a)}#3}\cref{Fig4}\crefformat{figure}, shows the data obtained when moving the top lens at an intermediate velocity. Again, it is not possible to fit the trajectory at intermediate distances when considering a purely viscous force with $b=4.2\textrm{~nm}$ (blue curve, corresponding to $A=0$), even if the discrepancy is less pronounced than at very low velocity. A fit can be achieved when including an electrostatic contribution (orange curve), with $v=1.13\pm0.04~\textrm{nm/s}$ and parameters for the static force which are consistent with the values extracted at very low velocities: $\lambda_\mathrm{s}=7.5\pm 0.5\textrm{~nm}$ and $A=0.18\pm 0.02\textrm{~mN/m}$.

In this regime, the viscous force is never negligible, whatever the distance considered. At intermediate distances, the viscous and electrostatic forces have the same orders of magnitude and contribute equally to the total force (\crefformat{figure}{Figure~#2#1{(b)}#3}\cref{Fig4}\crefformat{figure},).

We now return to the analyses made in previous studies \cite{Gebbie2013a,EspinosaMarzal2014a,EspinosaMarzal2014b,Cheng2015a,Gebbie2015a,Smith2016a}. In those studies, comparable measurements were performed with ionic liquids of similar or higher viscosity and using velocities close to $1~\textrm{nm/s}$. Viscous forces were not explicitly included in fitting the data; instead, in those experiments, the velocity was varied over a small range to ensure the force and decay was not velocity-dependent. This means that the motion imposed on the top lens \textcolor{black}{(i.e the piezo velocity)} was simply deduced from a linear fit of the trajectory at large distances. The previous fitting procedure appears particularly inappropriate when looking at the starting curvature of the trajectory, which we are now able to resolve clearly (inset of \crefformat{figure}{Figure~#2#1{(a)}#3}\cref{Fig4}\crefformat{figure},), which is a signature of the presence of a non-negligible viscous contribution. Nonetheless, we now make such a \textcolor{black}{linear} fit with our present data (dashed line in gray), to \textcolor{black}{quantify the error committed when not taking the viscous contribution into account}. The \textcolor{black}{fitted} velocity is $\textcolor{black}{v=~}0.99\pm0.04~\textrm{nm/s}$, slightly smaller but close to the velocity extracted with our full model. A force \textcolor{black}{is} then deduced from the difference between the trajectory and this base line (gray line in \crefformat{figure}{Figure~#2#1{(b)}#3}\cref{Fig4}\crefformat{figure}) \textcolor{black}{and is finally} fitted with a simple exponential law to extract a screening length. As we see in \crefformat{figure}{Figure~#2#1{(b)}#3}\cref{Fig4}\crefformat{figure}, it turns out that this ill-defined force is well superimposed on the electrostatic contribution obtained from our model (red curve). As a result, the same value for the screening length is given by the two methods.

This surprising matching of the electrostatic screening length, both with and without accounting for the viscous contribution, can be interpreted qualitatively by inspecting the two contributions to the total force on linear scales (inset of \crefformat{figure}{Figure~#2#1{(b)}#3}\cref{Fig4}\crefformat{figure}). At distances larger than $\sim 11~\mathrm{nm}$ (indicated by a vertical dashed line in gray), the viscous force is linear to very good approximation (except, of course, for the curvature at $D\gtrsim 60~\mathrm{nm}$). At distances shorter than $\sim 11~\mathrm{nm}$, the electrostatic part is bigger than the viscous one and varies more slowly, thus the departure from linearity of the viscous force is not substantial. We find, therefore, that calculating a force from a direct linear fit of the trajectory is an uncontrolled way to subtract the viscous contribution to the interaction. Neglecting the viscous effects turns out -- in this and similar cases -- to take them into account.

\begin{figure}[h!]
\includegraphics{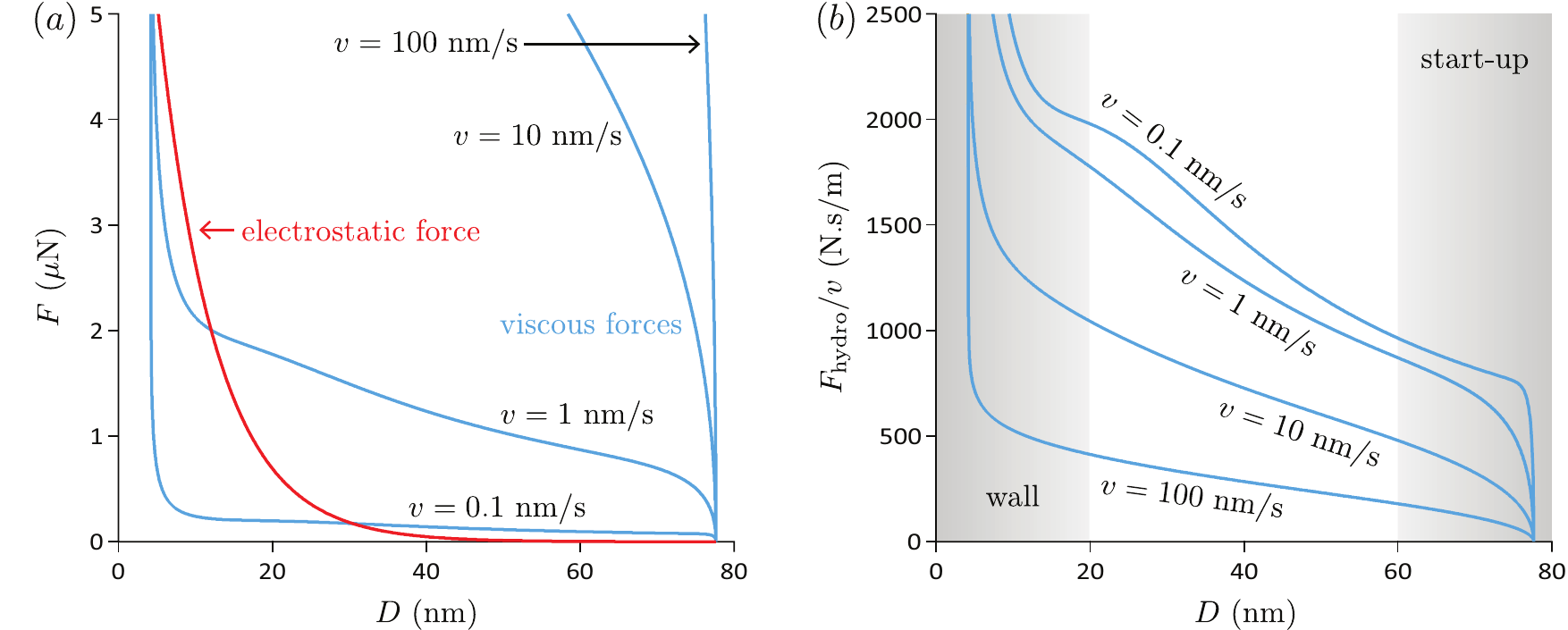}
\caption{\textcolor{black}{(a)~Electrostatic and viscous forces calculated numerically for four approaches at $v\in\{0.1~\mathrm{nm/s};1~\mathrm{nm/s};10~\mathrm{nm/s};100~\mathrm{nm/s}\}$, with $b=4.2\textrm{~nm}$, $\lambda_\mathrm{s}=7.5\pm0.5\textrm{~nm}$ and $A=0.18\pm0.02\textrm{~mN/m}$. (b)~Ratio of the viscous force to the approach velocity as a function of the distance between the surfaces, for the same trajectories.}}
\label{Fig5}
\end{figure}

\textcolor{black}{One can ask why the viscous force, that depends on~$\dot{D}/D$, is in fact found to be linear with~$D$. To discuss this point, we solved numerically the equation of motion for four approach velocities $v\in\{0.1~\mathrm{nm/s};1~\mathrm{nm/s};10~\mathrm{nm/s};100~\mathrm{nm/s}\}$ (the other parameters being kept constant as $b=4.2\textrm{~nm}$, $\lambda_\mathrm{s}=7.5\pm0.5\textrm{~nm}$ and $A=0.18\pm0.02\textrm{~mN/m}$), and we computed the corresponding electrostatic and viscous forces (\crefformat{figure}{Figure~#2#1{(a)}#3}\cref{Fig5}\crefformat{figure}~). To compare the dissipative forces over 3 decades of approach velocities, it is convenient to plot the ratio of the viscous force~$F_\mathrm{hydro}$ to the approach velocity~$v$, that can be interpreted as a friction coefficient (\crefformat{figure}{Figure~#2#1{(b)}#3}\cref{Fig5}\crefformat{figure}~). For all the trajectories, i.e. for the complete range of velocities explored experimentally, we observe a significant curvature at large distances caused by the sudden start-up of the piezo, a strong curvature at short distances due to the close vicinity of the solid surfaces, and a reasonably linear regime at intermediate distances. Such quasi-linear regime at intermediate distances is not intrinsic to the mathematical form of the viscous force, but results from the competition between the expansion of the piezo on the top lens that tends to accelerate the squeeze-out of the liquid and the compression of the spring on the bottom lens that tends to decelerate it. Interestingly, the "friction coefficient" is found to decrease with the velocity. This is because the higher the velocity, the greater the difference between the top lens velocity $v$ and the maximum relative velocity~$\dot{D}$ at intermediate distances. The two surfaces are in contact before a steady-state at~$\dot{D}=v$ can be reached, thus the viscous force~$F_\mathrm{hydro}$ increases slower than the velocity~$v$. Discrepancy from linearity is clearly visible for the lowest velocity, and is due to the electrostatic contribution that becomes dominant for $D\lesssim30~\mathrm{nm}$ (\crefformat{figure}{Figure~#2#1{(a)}#3}\cref{Fig5}\crefformat{figure}~). The effect of viscosity is tiny at very low velocity, and it is of course this regime that should be preferred to have the best determination of the static contribution.}

\section{Conclusion}
\label{Conclusion}

In this paper we tackled the question of the contribution of the viscous force to the long-range interactions previously observed for nanoconfined ionic liquids. Using $[\mathrm{C_2 C_1 Im}][\mathrm{NTf_2}]$ confined between mica surfaces in a SFB, we explored the effect of approach velocity over three orders of magnitude and down to $0.07~\mathrm{nm/s}$, and we systematically analyzed the data by solving numerically an equation of motion that combines viscous and electrostatic forces. This systematic study allowed us to disentangle the dynamic and static contributions to the interaction, leading to the following conclusions. First, no significant deviation from classical hydrodynamics is found for this nanoconfined ionic liquid at high velocity, and an excellent agreement is achieved when incorporating a negative slip boundary condition with a negative slip length $b=4.2\pm 0.7\textrm{~nm}$. In other words, a nanometric film of liquid in the vicinity of the solid surface does not participate in the flow. We note that this shift of the hydrodynamic zero is in the range of the structural force and corresponds to the confinement at which stick-slip behavior is typically observed in friction experiments. All these independent observations point toward the presence of ordered, solid-like layers of liquid molecules in the vicinity of the solid surface. Then, we showed that an equilibrium, exponentially decaying force has to be invoked in order to describe quantitatively the behavior at low velocity. The screening length $\lambda_\mathrm{s}=8.0\pm 0.5\textrm{~nm}$ extracted from our fitting procedure is in agreement with the values previously reported for this system. This appears surprising given that in previous studies the viscous force was not taken into account explicitly. We interpreted this surprising matching by looking at an approach at intermediate velocity, comparable to the velocities used in previous studies. We showed that analyzing the data as if there is no viscous force is an (uncontrolled) way to subtract the viscous contribution to the total force, and thus led to a correct determination of the screening length.

Such long-range electrostatic interactions have also been reported using colloid probe AFM \cite{Hjalmarsson2017a} . Similarly, one can ask what is the contribution of viscous forces in these observations. The ratio of the viscous force to the electrostatic force expressed in this paper varies as $R \dot{D}$. SFB typically uses a radius $\sim 1~\mathrm{cm}$ and a velocity $\sim 1~\mathrm{nm/s}$, whereas colloid probe AFM typically uses a radius $\sim 10~\mathrm{\mu m}$ and a velocity $\sim 100~\mathrm{nm/s}$. For a given system, the ratio of the viscous force to the electrostatic force for colloid probe AFM is typically ten times smaller than for SFB. Viscous effects are thus expected to have less impact in colloid probe AFM experiments although the difference is not large and similar viscous effects will often be significant in AFM studies. In general it is necessary to consider, for any scenario involving viscous drainage in combination with equilibrium surface forces, the contribution of each to the total measured force. 
In our study, all the trajectories recorded while exploring 3 orders of magnitude of approach velocity have been fitted quantitatively and consistently by a model that simply combines independent viscous and electrostatic forces. With the current parameters it was not found necessary to consider electrokinetic effects. It would be interesting to investigate more the possibility of such phenomena, for example by exploring higher velocities or by simultaneously applying electric fields across the liquid gap.

\section*{Acknowledgments}
\label{Acknowledgments}

We acknowledge Joelle Frechette and Yumo Wang for helping us to perform a comparative measurement with a PDMS melt. S.P. and R.L are supported The Leverhulme Trust (RPG-2015-328) and the ERC (under Starting Grant No. 676861, LIQUISWITCH). R.L is supported by the EPA Cephalosporin Junior Research Fellowship and Linacre College (University of Oxford). SP is grateful for research leave enabled by the Philip Leverhulme Prize.

%




\end{document}